\documentclass[conference,twocolumn]{IEEEtran}

\usepackage{xcolor}
\usepackage{soul}
\usepackage{multirow} 
\usepackage{cite}
\usepackage{relsize}
\usepackage[labelsep=period]{caption}
\usepackage{amsmath}
\usepackage{varwidth}
\usepackage{amssymb}
\usepackage{algorithm}
\usepackage{algpseudocode}
\usepackage{mathtools}
\usepackage[cmintegrals]{newtxmath}
\usepackage{array}
\usepackage[caption=false,font=normalsize,labelfont=sf,textfont=sf]{subfig}
\usepackage{fixltx2e}
\usepackage{stfloats}
\usepackage{enumitem}

\makeatletter
\newcommand{\distas}[1]{\mathbin{\overset{#1}{\kern\z@\sim}}}%
\newsavebox{\mybox}\newsavebox{\mysim}
\newcommand{\distras}[1]{%
  \savebox{\mybox}{\hbox{\kern3pt$\scriptstyle#1$\kern3pt}}%
  \savebox{\mysim}{\hbox{$\sim$}}%
  \mathbin{\overset{#1}{\kern\z@\resizebox{\wd\mybox}{\ht\mysim}{$\sim$}}}%
}
\makeatother
\ifCLASSINFOpdf
  \usepackage[pdftex]{graphicx}
  \graphicspath{{./SmartGridComm/figs/}}
\else
\fi

\begin{document}

\title{\huge{Demand Responsive Dynamic Pricing Framework for Prosumer Dominated Microgrids using Multiagent Reinforcement Learning }}

% \IEEEauthorblockA{\textit{EECS Department} \\
% \textit{University of Kansas}\\
% Lawrence KS, USA \\
% Email}
% \and
% \IEEEauthorblockA{\textit{EECS Department} \\
% \textit{University of Kansas}\\
% Lawrence KS, USA \\
% Email}
% \and
% \IEEEauthorblockA{\textit{EECS Department} \\
% \textit{University of Kansas}\\
% Lawrence KS, USA \\
% Email}
% \and
% \IEEEauthorblockA{\textit{EECS Department} \\
% \textit{University of Kansas}\\
% Lawrence KS, USA \\
% Email}
% \and
% \IEEEauthorblockA{\textit{EECS Department} \\
% \textit{University of Kansas}\\
% Lawrence KS, USA \\
% Email}
% \and
% \IEEEauthorblockA{\textit{EECS Department} \\
% \textit{University of Kansas}\\
% Lawrence KS, USA \\
% Email}
% }

{\author{\IEEEauthorblockN{Amin Shojaeighadikolaei, Arman Ghasemi, Kailani R. Jones, \\Alexandru G. Bardas, Morteza Hashemi, Reza Ahmadi}}}
% \IEEEauthorblockA{Department of Electrical Engineering and Computer Science, University of Kansas, Lawrence, KS, USA} 
% E-mails: \{amin.shojaei, arman.ghasemi, kailanij, alexbardas, mhashemi, ahmadi\}@ku.edu}}

\maketitle

\begin{abstract}
Demand Response (DR) has a widely recognized potential for improving grid stability and reliability while reducing customers' energy bills.  However, the conventional DR techniques come with several shortcomings, such as inability to handle operational uncertainties and incurring customer disutility, impeding their wide spread adoption in real-world applications.  This paper proposes a new multiagent Reinforcement Learning (RL) based decision-making environment for  implementing a Real-Time Pricing (RTP) DR technique in a prosumer dominated microgrid. The proposed technique addresses several shortcomings common to traditional DR methods and provides significant economic benefits to the grid operator and prosumers.  
To show its better efficacy, the proposed DR method is compared to a baseline traditional operation scenario in a small-scale microgrid system.  Finally, investigations on the use of prosumers' energy storage capacity in this microgrid highlight the advantages of the proposed method in establishing a balanced market setup.             

\end{abstract}

\IEEEpeerreviewmaketitle


\section{Introduction}
Demand-side management, also known as Demand Response (DR), is one of the most widely studied topics in the context of the smart grid~\cite{SIANO2014461}. In order to flatten the demand curve, conventional time-based DR methods such as the Real-Time Pricing (RTP) approach~\cite{HAIDER2016166} rely on dynamically changing the electricity price to motivate the customers to alter their energy use profile~\cite{6861959}. This potentially improves the grid stability and reliability by shifting the peak demand and decreasing the need for peaking power plants while offering reduced energy bills to residential customers~\cite{BookIR,9000240}.  

Nevertheless, the conventional time-based DR methods typically assume that the pricing policies are deterministic and decided ahead of time~\cite{7152979,6102349,9042589}, or consider that pricing policies follow a random process with known properties~\cite{SHOREH201631}. Thus, conventional DR methods are either not able to offer convergence to the optimal solution in presence of uncertainties in the environment or the mathematical formulations become cumbersome to model uncertainties~\cite{7938652}, which makes them unsuitable for real-world implementations.  

Moreover, the conventional time-based DR approaches almost exclusively focus on altering the preferred electricity consumption pattern of the customers, e.g., by changing the temperature set point on air conditioning systems or delaying the use of major appliances, in order to shift their load to off-peak periods. These approaches often lead to customer dissatisfaction (disutility), and, as a consequence, have not been widely adopted. 
Psychologically, the consumers value their comfort much higher than economic savings provided by traditionally proposed DR approaches~\cite{7024930}.  

Integration of energy storage into residential photovoltaic (PV) systems, EV and PV forecasting should provide energy consumers \& producers, commonly known as \emph{prosumers}, with more flexibility to participate in DR programs while minimizing their disutility\cite{8662200,8786152,PANAMTASH2020336}. In other words, prosumers should be able to shape their demand profile in real-time regardless of their consumption profile~\cite{YANG2016353}.  As a result, prosumers can potentially receive greater economic benefits by selling their excess energy to the grid at higher prices, while aiding grid support services~\cite{8258959}. 
Achieving the aforementioned benefits demands for a novel DR approach on the prosumers' side. This DR approach considers various factors such as state-of-charge (SOC) of the storage device, household consumption profile, and real-time electricity price and PV generation levels. Furthermore, taking advantage of the flexibility provided by prosumers requires a modern grid management strategy which can treat the storage capacity of households as a grid asset that can be dispatched by properly incentivizing the households for DR participation, and leverage this asset for grid cost and performance optimization.

This paper proposes a multiagent deep Reinforcement Learning (RL) framework for implementing a new RTP-based DR technique in a prosumer microgrid that provides both, prosumers and the grid operator, with the aforementioned flexibility and greater economic benefits.  The main contributions of the proposed framework are summarized as follows: 
a) {\it Real-time learning}: Grid and prosumer agents learn the optimal price and DR participation policy by interacting with the environment in real-time, rather than using a complex grid dynamic model for optimization.  Therefore, the proposed method is applicable to the high dimensional and non-stationary environment of the grid with much less computational burden than traditional DR methods, allowing for real-world implementations; 
% b) {\it Adaptive against environment variations}: The proposed method converges to an optimal solution even in the presence of typical uncertainties in a real-world system; 
b) {\it Altering the households' grid injection patterns}: The goal of the proposed prosumer-side DR algorithm is {\it not} to alter the consumption pattern of households which can typically lead to customer dissatisfaction. Rather, the prosumer-side DR algorithm provides cost savings by altering the households' grid injection pattern, using the flexibility provided by the energy storage and PV generation; and 
c) {\it Balanced market}: The proposed framework makes better use of prosumers' energy storage capacity, allowing for significant prosumer electricity bill reduction, as well as considerable grid economic benefit improvement, with a reasonably sized battery pack. Our results also show that after a certain threshold value, a larger battery size on the prosumer side does not necessarily yield a much higher profit. This facilitates a balanced market setup where trying to abruptly and unilaterally manipulate the pricing scheme is not in anyone's financial interest. 

\textbf{Related Work:} In recent years there has been a growing interest in application of RL to the problem of dynamic pricing and DR.  A comprehensive survey of published works in this area is provided in~\cite{VAZQUEZCANTELI20191072}. Among the surveyed publications, the works in~\cite{DynamicPrice} and~\cite{11Dynamic} are closely aligned with our work.  The authors in \cite{DynamicPrice} propose a RL-based dynamic pricing and energy consumption scheduling framework that can work without priori information and leads to reduced system costs.  The proposed framework targets regular customers without grid injection capability and assumes that customer behavior is myopic and deterministic, i.e. each costumer is trying to minimize its cost in every single time slot. In contrast, our work takes advantage of the PV generation and storage capacity of prosumers, leading to greater flexibility for DR participation, and enables the prosumers to make decisions that lead to long-term optimization of their accumulative economic benefit, rather than minimizing their instantaneous cost.  On the other hand, the authors in \cite{11Dynamic} present a RL-based dynamic pricing algorithm which can promote service provider profitability and reduce energy costs for the customers.  However, similar to~\cite{DynamicPrice}, this work only deals with regular electricity consumers, rather than prosumers with generation capability.  A multi-Agent using deep reinforcement learning for a distributed energy resources in Smart Grids also provided in \cite{SGC} which is highly related to our work.  

The reminder of this paper is structured as follows: Section~\ref{sec:model} details the proposed RL based DR approach.  Section~\ref{sec:simulations} implements the proposed method for a small-scale microgrid as a case study and provides simulation results to verify efficacy of the proposed method. Finally, our concluding thoughts are presented in Section~\ref{sec:conclusions}.

\section{Proposed RL Based DR Method}\label{sec:model}

\begin{figure*}
\centering
    \includegraphics[scale=0.58, trim =0.5cm .5cm 0.5cm 0.5cm, clip]{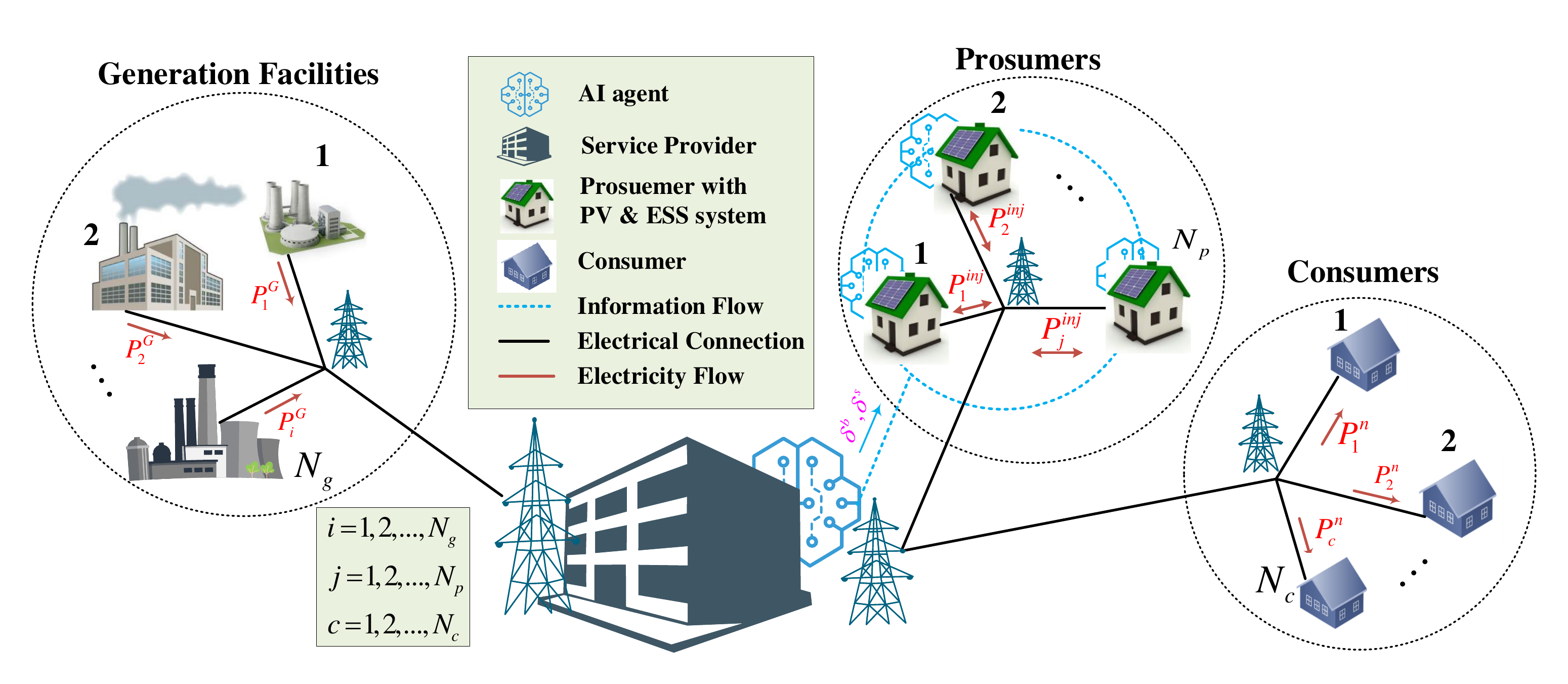}
    \caption{Small scale microgrid system that consists of generation facilities, traditional consumers, and prosumers. Grid and prosumers are equipped with reinforcement learning (RL) agents to dynamically adjust their policies in terms of buy/sell prices, and power injection.  
    }
    \label{singleLine}
    \vspace{-.15in}
\end{figure*}

Fig.~\ref{singleLine} illustrates the envisioned microgrid system with conventional generation facilites, prosumers, and consumers.  Prosumers are entities that can produce energy locally from renewable resources, store the excess energy in their battery, sell energy into the grid, or buy electricity from the grid for local consumption. The prosumers can make a profit by selling electricity to the grid at a dynamic \$/kWh price of ($\delta^b$), referred to as buy price hereinafter.  On the other hand, the prosumers incur a cost when buying electricity from the grid at a \$/kWh price referred to as sell price ($\delta^s$), hereinafter.  The goal of prosumer agents is to maximize its long term profit by determining an optimal charge/discharge policy for their energy storage devices.  Similarly, the grid can buy energy from prosumers at a price of ($\delta^b$) and incur a cost, or sell electricity to prosumers at a price of ($\delta^s$) to make a profit.  The goal of the grid agent is defined as maximizing the long term profit of the grid by determining an optimal buy price ($\delta^b$) policy. Therefore, we define the following optimization problem for grid agents,       
% The proposed electricity market model is shown in Fig.~\ref{singleLine}  As pictured, this model encompasses a grid agent (GA) and several prosumer agents (PAs).  The learning environment is the combination of governing equations of the grid and prosumer’s physical systems, the operational limitations of the power grid and the prosumers, and external factors such as the time of day or PV generation level as explained in the physical model subsection below.  The states, actions, and rewards for the agents are also explained further down in the learning system model.
% \begin{figure}[hbt]
%   \centering
%     \includegraphics[scale=0.4]{SmartGridComm2020-paper/figs/Microgrid.png}
%     \caption{Microgrid Model}
%     \label{Microgrid}
% \end{figure}
% In a power system with $n_g$ generators each with power output level of $\ P_{G_i}$ where $i\in\left\{1,\ldots,n_g\right\}$ , and $n_h$ prosumers each with power injection level of $P_{H_j}$ where $j\in\left\{1,\ldots,n_h\right\}$, and the total grid demand $P_D$, 
% \begin{align}\label{eq:1}
%     &\textrm{Maximize}\,\,\,\,\,
%     \begin{aligned}[t]
%       & {{\rm M}\left( T \right) = V\left( {{P^{dm}}\left( T \right)} \right) - \\
%       & \sum\limits_{i = 1}^{{N_g}} {{F_i}\left( {P_i^G\left( T \right)} \right) - \sum\limits_{j = 1}^{{N_p}} {{C_j}} } \left( {P_j^{inj}\left( T \right)} \right)}
%     \end{aligned} \\
%     &\text{subject to:} \notag \\
%     & {\sum\limits_{q = 1}^{{N_c} + {N_p}} {P_q^{d}\left( t \right) = \sum\limits_{i = 1}^{{N_g}} {P_i^G\left( t \right)  + \sum\limits_{j = 1}^{{N_p}} {P_j^{inj}\left( t \right)} } } } \ ,\label{eq:2} \\
%     & {P_i^{G,\min}\ (t) \le P_i^G \ (t) \le P_i^{G,\max}\ (t)} \ 
% \end{align}
 \begin{align} 
 \begin{cases}
 \mathop{\mathrm{maximize}} & \text{M} \left( T \right) \\ 
 \text{subject to:} &  \mathlarger{\sum}\limits_{q = 1}^{\small{N_c+N_p}} P_q^{d}\left(t\right) = \mathlarger{\sum}\limits_{i = 1}^{{N_g}} P_i^G\left( t \right)  + \mathlarger{\sum}\limits_{j = 1}^{N_p} {P_j^{inj}\left( t \right)}   \\
 & ~ P_i^{G, \text{min}} (t) \le P_i^G (t) \le P_i^{G, \text{max}} (t),  
 \end{cases}
 \label{eq:1}
 \end{align}
where $\text{M}(T)$ is the grid profit defined as follows: 
$$
\text{M} \left( T \right) = V\left( {P^{dm}}\left( T \right) \right) - \sum\limits_{i = 1}^{N_g} {F_i}\left( {P_i^G\left( T \right)} \right) - \sum\limits_{j = 1}^{{N_p}} {{C_j}} \left( {P_j^{inj}\left( T \right)} \right). 
$$
In addition, we have: 
\begin{align}\label{eq:4}
V\left( {{P^{dm}}\left( T \right)} \right) = \int_0^T {{P^{dm}}\left( t \right)} \,\,{\delta ^s}\left( t \right)dt\ ,
\end{align}
\begin{align}\label{eq:5}
{C_j}\left( {P_j^{inj}\left( T \right)} \right) = 
% \left\{ \begin{gathered}
%   \int\limits_0^T {P_j^{inj}\left( t \right){\delta ^b}\left( t \right)dt}\,\,\,\textrm{for}\,P_j^{inj}(t) > 0 \hfill \\
%   0\,\,\,\,\,\,\,\,\,\,\,\,\,\,\,\,\,\,\,\,\,\,\,\,\,\,\,\,\,\,\,\,\,\,\,\,\,\,\,\,\,\,\,\,\,\,\,\,\,\,\textrm{for}\,P_j^{inj}(t) \leqslant 0\, \hfill \\ 
% \end{gathered}  \right.\ ,
 \begin{cases}
  \int\limits_0^T {P_j^{inj}\left( t \right){\delta ^b}\left( t \right)dt} & \textrm{for}\  P_j^{inj}(t) > 0  \\
  0 & \textrm{for}\ P_j^{inj}(t) \leq 0.
\end{cases}
\end{align}
In which  $V({P^{dm}}(T))$ represents the accumulative grid revenue by selling electricity to the households in time horizon $T$, ${{C_j}( {P_j^{inj}( T )})}$ represents the accumulative grid cost of buying excess energy from $j^{th}$ prosumer in time horizon $T$, ${{F_i}( {P_i^G(T)})}$ represents the accumulative cost of buying electricity form $i^{th}$ generation facility time horizon $T$, $N_c$ , $N_g$ and $N_p$ are the number of consumers, generation facilities and prosumers respectively. The first equality constraint 
in \eqref{eq:1} 
represents the grid's power balance requirement which needs to be maintained at all times, $P^{dm}$ is the total demand of the network, ${P_q^d\left( t \right)}$ is the demand of household $q$ , and $P_j^{inj}$ is the power injection into the grid by $j^{th}$ prosumer. 

Similarly we define the following optimization problem for $j^{th}$ prosumer, 
% \begin{align}\label{eq:6}
%     &\textrm{Maximize}\,\,\,\,\,
%     \begin{aligned}[t]
%       & {{\rm U}_j}\left( T \right) = V_j^p\left( {P_j^{inj}\left( T \right)} \right) - C_j^p\left( {P_j^{inj}\left( T \right)} \right)
%     \end{aligned} \ , \\
%     &\text{subject to} \notag \\
%     & P_j^{inj}\left( t \right) + P_j^{batt}\left( t \right) + P_j^{c}\left( t \right) = P_j^{pv}\left( t \right)\label{eq:7} \ , \\
%     & \left| {P_j^{inj}\left( t \right)} \right| \le P_j^{inj,\max }\label{eq:8} \ ,  \\
%     & \left| {P_j^{batt}\left( t \right)} \right| \le P_j^{batt,\max }\label{eq:9} \ ,  \\
%     & 0 \le P_j^{pv}\left( t \right) \le P_j^{pv,\max }\label{eq:10} \ , \\
%     & SoC_j^{\min } \le So{C_j}\left( t \right) \le SoC_j^{\max }\label{eq:11} \ ,
% \end{align}
 \begin{align} 
 \begin{cases}
 \mathop{\mathrm{maximize}} & {{\rm U}_j}\left( T \right) \\ 
 \text{subject to:} &  P_j^{inj}\left( t \right) + P_j^{batt}\left( t \right) + P_j^{c}\left( t \right) = P_j^{pv}\left( t \right)  \\
 & \left| {P_j^{inj}\left( t \right)} \right| \le P_j^{inj,\max } \\
 & \left| {P_j^{batt}\left( t \right)} \right| \le P_j^{batt,\max } \\
 & 0 \le P_j^{pv}\left( t \right) \le P_j^{pv,\max } \\
 & SoC_j^{\min } \le So{C_j}\left( t \right) \le SoC_j^{\max }, 
 \end{cases}
 \label{eq:6}
 \end{align}
where ${{\rm U}_j}(T) = V_j^p\left( {P_j^{inj}\left( T \right)} \right) - C_j^p\left( {P_j^{inj}\left( T \right)} \right)$ is the $j^{th}$ prosumer accumulative profit in time horizon $T$, $V_j^p(P_j^{inj}(T))$ represents the $j^{th}$ prosumer accumulative revenue for selling excess electricity to the grid in time horizon $T$ calculated using the same equation as \eqref{eq:5} considering $V_j^p(P_j^{inj}(T))=C_j(P_j^{inj}(T))$, and $C_j^p(P_j^{inj}(T))$ represents the $j^{th}$ prosumer accumulative cost of buying electricity from the grid in time horizon $T$ calculated as,
% \begin{align}\label{eq:12}
% {C_j^p}\left( {P_j^{inj}\left( T \right)} \right) = \left\{ \begin{gathered}
%   \int\limits_0^T {P_j^{inj}\left( t \right)\,.\,{\delta ^s}\left( t \right)dt}\,\,\,\textrm{for}\,P_j^{inj}(t) < 0 \hfill \\
%   0\,\,\,\,\,\,\,\,\,\,\,\,\,\,\,\,\,\,\,\,\,\,\,\,\,\,\,\,\,\,\,\,\,\,\,\,\,\,\,\,\,\,\,\,\,\,\,\,\textrm{for}\,{P_j^{inj}\left(t \right) \ge 0} \\ 
% \end{gathered}  \right.\ ,
% \end{align}
\begin{align}\label{eq:12}
{C_j^p}\left( {P_j^{inj}\left( T \right)} \right) = 
\begin{cases}
  \int\limits_0^T {P_j^{inj}\left( t \right){\delta ^s}\left( t \right)dt} & \textrm{for}\,P_j^{inj}(t) < 0, \\
  0 & \textrm{for}\,{P_j^{inj}\left(t \right) \ge 0}.   
\end{cases}
\end{align}
$P_j^{pv}(t)$ is the PV generation with the peak generation of $P_j^{pv,\max }$ for $j^{th}$ household, $P_j^{c}(t)$ is consumption for $j^{th}$ household, $P_j^{inj,\max }$ represents the maximum allowable power injection for $j^{th}$ household, $P_j^{batt}(t)$ represents the energy storage charge/discharge power with maximum allowable charge/discharge power of $P_j^{batt,\max }$ for $j^{th}$ household, $So{C_j}\left( t \right)$ is the state of charge for $j^{th}$ prosumer where $SoC_j^{\min }$ and $SoC_j^{\max }$ are minimum and maximum allowable state of charge. The state of charge is calculated as,
\begin{align}\label{eq:13}
So{C_j}\left( t \right) = So{C_j}\left( 0 \right) + \frac{1}{{\eta _j}}\int_0^t {P_j^{batt}\left( \tau \right)\,d\tau}, 
\end{align}
where $So{C_j}\left( 0 \right)$ is the initial state of charge and ${\eta _j}$ is the energy storage capacity of $j^{th}$ prosumer.

\begin{figure}
   \centering
    \includegraphics[scale=0.6, trim =0.5cm .5cm 0.5cm 0.5cm, clip]{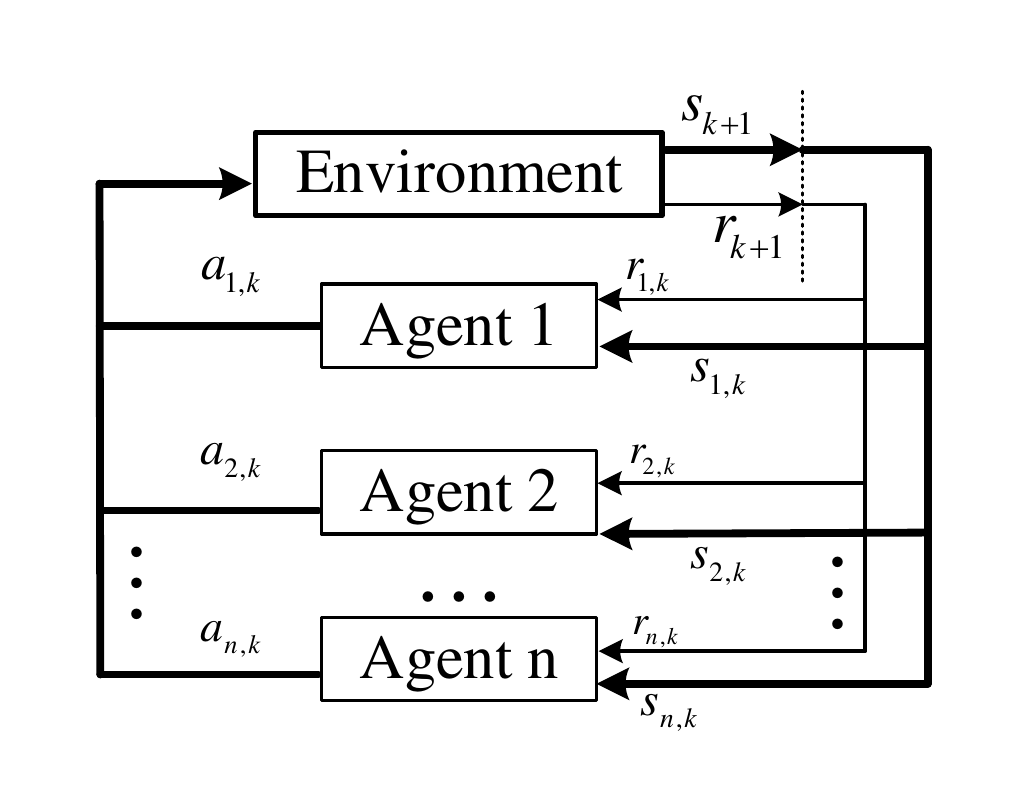}
    \vspace{-.1in}
    \caption{A typical reinforcement learning framework wherein agents learn their optimal actions through observing the environment states, taking actions, and receiving rewards. }
    \label{reinforcement}
    \vspace{-.15in}
\end{figure}

Each agent uses RL for solving the defined optimization problems in real-time.  The agents interact with the environment in order to maximize a specified reward. The agents receive positive rewards for taking desirable actions and negative rewards for taking undesirable actions. This learning process is modeled as a Markov Decision Process (MDP) for a Multi-agent Reinforcement Learning (MARL) environment  defined by a tuple of $\{ {\rm{N,\ S,\ \{ }}{{\rm{A}}_{\rm{i}}}{\rm{\} ,\ P, \ \{ }}{{\rm{R}}_{\rm{i}}}{\rm{\} , \ \gamma }}\} $ where $\rm{N}$ is the set of agents, $\rm{S}$ is the set of states of the environment, ${\rm{A}}_{\rm{i}}$ is the set of actions for $i^{th}$ agent, $\textrm{P}$ is the set of transition function, ${\rm{R}}_{\rm{i}}$ is the immediate reward function set for $i^{th}$ agent, and ${\rm{\gamma }} \in \left[ {0,1} \right]$ is the discount factor. The agent-environment interaction for an MDP is shown in Fig~\ref{reinforcement}. At each discretized time index $k$, the agent $\rm{N_i}$ selects an action ${a_{i,k}} \in {{\rm{A}}_{\rm{i}}}$ based on observing the current state of the environment denoted by $s_k$. Subsequently, the agent receives a numerical feedback signal known as reward, ${r_{i,k}} \in {{\rm{R}}_{\rm{i}}}$, and transitions to a new state $s_{k+1}$.  In the proposed framework, each agent takes actions only based on its own local. In other word, each agent can only observe partial features of the entire environment. Thus, the grid agent can observe the following environment states, 
\begin{align}\label{eq:14}
{s_{1,k}} = \left\{ {{{\mathbf{F}}_k}, \ {{\mathbf{\Omega }}_k},\ P_{1,k}^{dm}} \right\} \in \textrm{S} \ ,
\end{align}
where ${{\mathbf{F}}_k}={\left[ {\begin{array}{*{20}{c}}
  {{F_{1,k}}}&{...}&{{F_{i,k}}} 
\end{array}} \right]^T}$ and ${{\mathbf{\Omega }}_k}={\left[ {\begin{array}{*{20}{c}}
  {{C_{1,k}}}&{...}&{{C_{j,k}}} 
\end{array}} \right]^T}$ are the vectors of the grid cost for buying electricity from the generation facilities and prosumers at time slot k respectively, ${{P}}_{1,k}^{dm}$ represents the total grid demand at time slot k.

Similarly, each prosumer agent can observe the following environment states,
\begin{align}\label{eq:15}
    & \begin{aligned}[t]
       & {s_{n,k}} = \left\{ {SoC_{n,k}, \ P_{n,k}^{pv},\ P_{n,k}^{c}, \ \delta _k^b} \right\} \in \textrm{S} \notag
    \end{aligned} \\
    & \textrm{for}\,\,\,n = 2,3, ... , {N_p} + 1 \ , 
\end{align}
where ${SoC_{n,k}}$ is the state of charge  of energy storage device of $n^{th}$ prosumer at time slot k, $P_{n,k}^{pv}$ and $P_{n,k}^{c}$ are the PV generation and local consumption of $n^{th}$ prosumer at time slot k, $\delta _k^b$ is the buy price at time slot k.

In this work, the grid agent controls the buy price to incentivize the DR participation of the prosumers and maximize its own reward.  Therefore, the buy price is the action of the grid agent denoted by ${a_{1,k}} = \delta _k^b \in {A_1}$.  On the other hand, the prosumer agents control the household's energy storage charge/discharge state in response to the dynamic buy price changes.  Therefore, the action for $n^{th}$ prosumer's agent is defined as ${a_{n,k}} \in {A_n}$ which denotes the charge/discharge command to the energy storage. 

Finally, the immediate reward function for the grid and prosumer agents are defined as,
\begin{align}\label{eq:16}
{r_{1,k}} = P_{1,k}^{dm} \times \delta _k^s - \sum\limits_{i = 1}^{{N_g}} {{F_{i,k}}\left( {P_{i,k}^G} \right)}  - \sum\limits_{j = 1}^{{N_p}} {P_{j,k}^{inj} \times \delta _k^b} \,\,\,\textrm{for}\,P_{j,k}^{inj} > 0 \ ,
\end{align}
\begin{align}\label{eq:17}
{r_{n,k}} = \left( \rho  \right)P_{n,k}^{inj} \times \delta _k^b + \left( {\rho  - 1} \right)P_{n,k}^{inj} \times \delta _k^s \ ,
\end{align}
where $\rho  \in \left\{ {0,1} \right\}$ in which $\rho  = 0$ when $P_{n,k}^{inj} \leqslant 0$ and $\rho  = 1$ when $P_{n,k}^{inj} > 0$.

According to the above terminology, the MDP trajectories for the grid agent and $n^{th}$ prosumer's agent begin with ${s_{1,1}},\ {a_{1,1}},\ {r_{1,2}}, \ {s_{1,2}}, \ {a_{1,2}}, \ {r_{1,3}},...$ and ${s_{n,1}}, \ {a_{n,1}}, \ {r_{n,2}},\ {  s_{n,2}}, \ {a_{n,2}}, \ {r_{n,3}},...$, respectively. The primary goal of each agent is to maximize the accumulative reward sequence formulated by ${G_1} = \mathlarger{\sum}\limits_{t = 0}^\infty  {{{\left( {{\gamma _1}} \right)}^t}}\,{r_{1,k + t + 1}}$ and ${G_n} = \mathlarger{\sum}\limits_{t = 0}^\infty  {{{\left( {{\gamma _n}} \right)}^t}} .\,{r_{n,k + t + 1}}$ where $0 \leqslant {\gamma _1} \leqslant 1$ and $0 \leqslant {\gamma _n} \leqslant 1$ are discount factors and $n = 2,3,...,{N_p} + 1$.  In this work, to find the optimal policy, Deep Q-Network (DQN) \cite{naderializadeh2019energy} is deployed to approximate the optimal action-value function which satisfies the Bellman Equation as,

\begin{align}\label{eq:18}
{{\hat Q}_{1,k + 1}}\left( {{s_{1,k}},{a_{1,k}}} \right) = \left( {1 - {\alpha _1}} \right){{\hat Q}_{1,k}}\left( {{s_{1,k}},{a_{1,k}}} \right) + & \nonumber \\ {\alpha _1}\left\{ {{r_{1,k}}\left( {{s_{1,k}},{a_{1,k}}} \right) + {\gamma _1}\mathop {\max }\limits_{{a_{1,k + 1}}} {{\hat Q}_{1,k}}\left( {{s_{1,k + 1}},{a_{1,k + 1}}} \right)} \right\}\ ,
\end{align}

\begin{align}\label{eq:19}
{{\hat Q}_{n,k + 1}}\left( {{s_{n,k}},{a_{n,k}}} \right) = \left( {1 - {\alpha _n}} \right){{\hat Q}_{n,k}}\left( {{s_{n,k}},{a_{n,k}}} \right) + & \nonumber \\ {\alpha _n}\left\{ {{r_{n,k}}\left( {{s_{n,k}},{a_{n,k}}} \right) + {\gamma _n}\mathop {\max }\limits_{{a_{n,k + 1}}} {{\hat Q}_{n,k}}\left( {{s_{n,k + 1}},{a_{n,k + 1}}} \right)} \right\}\ ,
\end{align}
where ${\hat Q}$ is the approximation of $Q$ that is estimated by a deep neural network. In this work, we use the updating mechanism provided in~\cite{SuttonBook} for Q-values.

\section{Case Study and Numerical Results}~\label{sec:simulations}

\begin{figure}[t]
   \centering
    \includegraphics[scale=0.39, trim =0.3cm .5cm 0.5cm 0.5cm, clip]{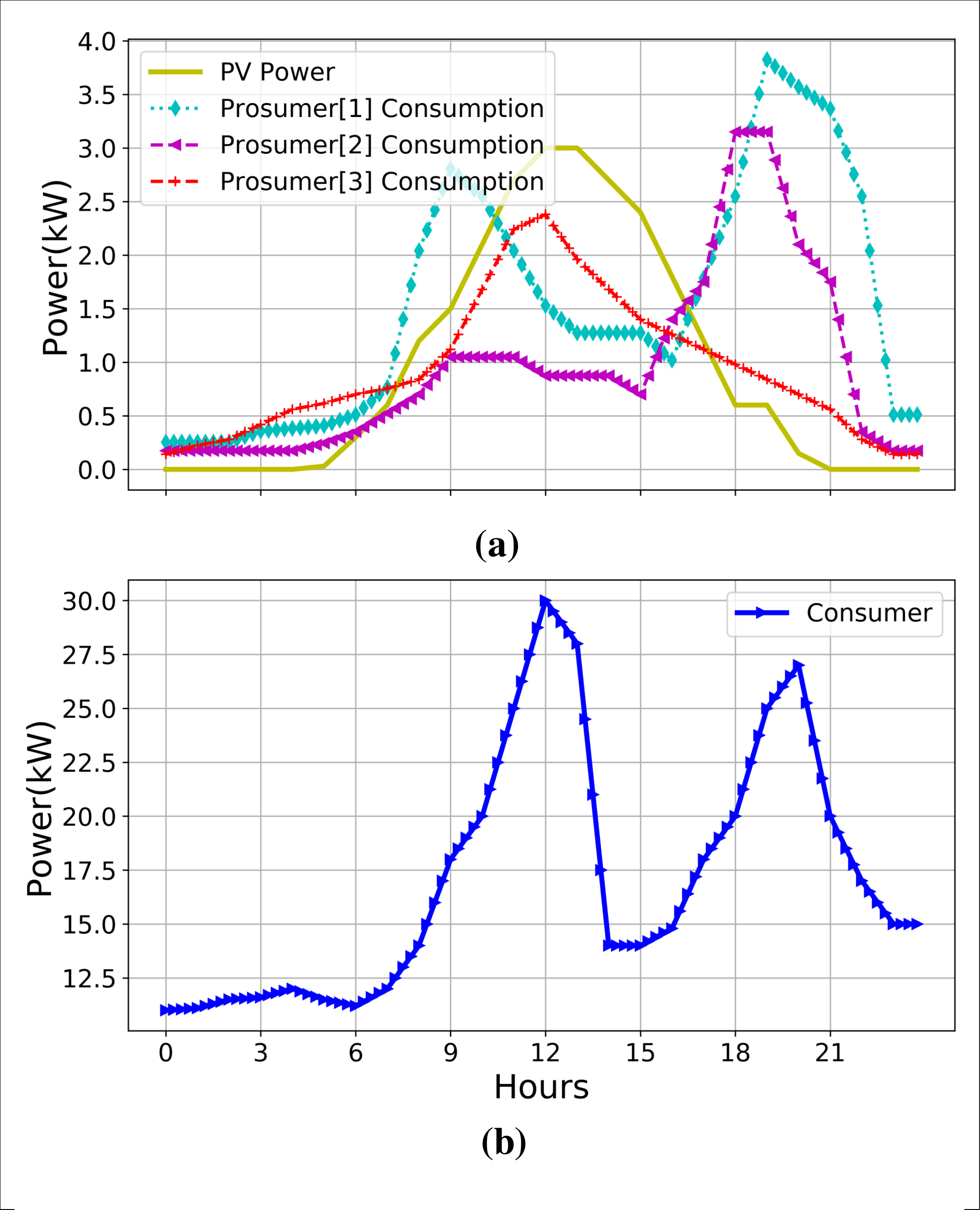}
    \caption{generation and consumption waveform sample (a) Generation and prosumers' consumption waveform (b) Consumer consumption waveform}
    \label{waveform}
    \vspace{-.2in}
\end{figure}

The proposed DR scheme is implemented on a small-scale microgrid similar to Fig.~\ref{singleLine}. The case study system is comprised of $(N_p=3)$ prosumers equipped with solar rooftop panels, energy storage system and smart agent, a conventional consumer representative of non-generational consumer $(N_c=1)$, and two generation facilities $(N_g=2)$ where one acts as a baseline generation facility and another acts as a reserve generation capacity. The prosumers' details are provided in Table.~\ref{SimulationParam}. Employed generation and consumption profiles for the prosumers are provided in Fig.~\ref{waveform}, where the generation and consumption profiles are derived from California ISO~\cite{CaliIso}.

% \mycomment{?}For the prosumers, we try to change the consumption profiles to show that the DR model works appropriately in different conditions.

Two scenarios have been simulated for analyzing the efficacy of the proposed DR scheme. In the first scenario, referred to as conventional scenario hereinafter, no DR scheme is applied to the microgrid.  The prosumers simply inject power to grid when there is an excess energy generation in the household and the battery capacity is full.  For the second scenario, the proposed DR scheme is implemented on the microgrid system and results are compared with the first scenario to evalute the effectiveness of the proposed method.

\begin{table}[b]
 \resizebox{\columnwidth}{!}{
    \begin{tabular}{c|c|c|c|c}
      \hline
      \multirow{2}{*}{\textbf{Item}} & \textbf{Max PV}&\textbf{ESS}&\textbf{Max}&\multirow{2}{*}{\textbf{Agent}} \\
      & \textbf{Generation}&\textbf{Capacity}&\textbf{Charge/Discharge}&\\
      \hline
      {Prosumer 1} & 4 kW & 6 kW & 2/-2 kW & Agent1 \\
      {Prosumer 2} & 4 kW & 12 kW & 2/-2 kW & Agent2 \\
      {Prosumer 3} & 4 kW & 8 kW & 2/-2 kW  & Agent3 \\
      \hline
    \end{tabular}
    }
    \caption{Microgrid Details}
    \label{SimulationParam}
    \vspace{-.15in}
\end{table}

% \begin{table*}[t!]
%   \begin{center}
%     \caption{Simulation parameters}
%     \label{SimulationParam}
%     \begin{tabular}{l|c|l}
%       \hline
%       \textbf{Parameter} & \textbf{Description} & \textbf{Value}\\
%       \hline
%       $P_{p{v_j}}^{\max }$ & Maximum PV Generation & [2-2.5] kW \\
%       $P_{b_j}^{\max }$ & Maximum allowable charge/discharge & 2/-2 kW \\
%       $P_{H_j}^{\max }$ & Maximum allowable power injection & 10 kW \\
%       $\phi_j^{\max }$ & Maximum state of charge & $0.9 \times {C_{b_j}}$ \\
%       $\phi_j^{\min }$ & Minimum state of charge & $0.1 \times {C_{b_j}}$ \\
%       $C_{b_j}$ & Energy storage capacity & [8-10] kW \\
%       ${\phi _j}(0)$ & Initial state of charge & [3-4] kWh \\
%       $\rho _c$ & [Sell price before 11:00am,sell price after 11:00am] & [0.05,0.095] \$/kWh \\
%       $\rho _b^t$ & Buy price for agentbased scenario & ${\{0.05,0.06,0.07,0.08,0.09,0.1}\}$\$/kWh \\
%       $\rho _b^t$ & Buy price for conventional scenario & 0.05 \$/kWh \\
%       $\left[ {P_{{G_1}}^{\min },P_{{G_1}}^{\max }} \right]$ & Limitation of base generation facility & [5,20] kW  \\
%       $\left[ {P_{{G_2}}^{\min },P_{{G_2}}^{\max }} \right]$ & Limitation of reserve generation facility & [0,50] kW  \\
%       \hline
%     \end{tabular}
%   \end{center}
% \end{table*}

\begin{figure}[t]
  \centering
    \includegraphics[scale=0.54, trim =0.5cm .07cm 0.05cm 0.3cm, clip]{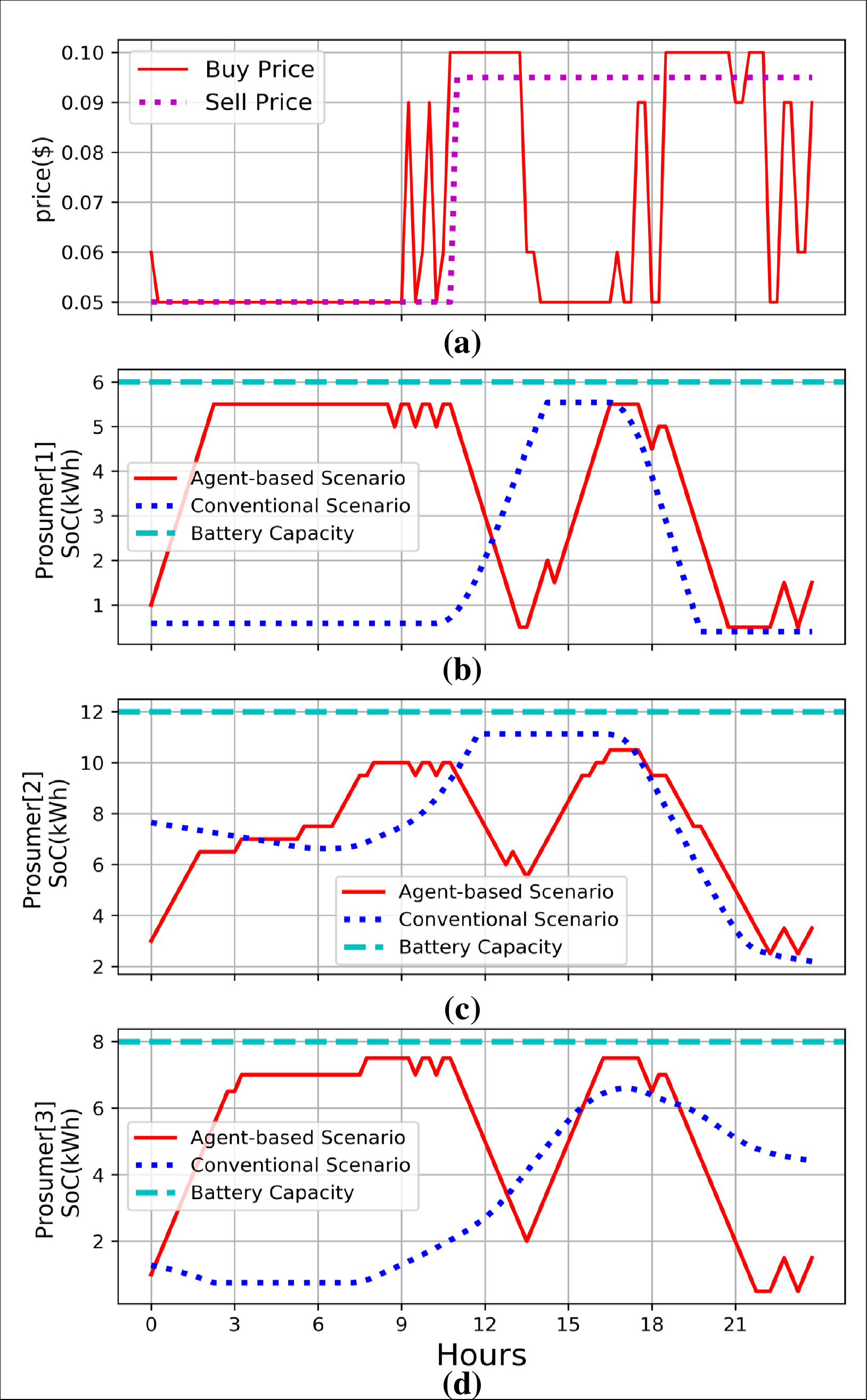}
    \caption{Simulation results after learning algorithm convergence for a day (24 hours) (a) Grid buying and selling price (b) Prosumer 1  battery SoC (c) Prosumer 2  battery SoC (d) Prosumer 3  battery SoC }
    \label{time-domain}
    \vspace{-.2in}
\end{figure}

\begin{figure}[t]
   \centering
    \includegraphics[scale=0.22, trim =0.5cm .15cm 0.5cm 0.5cm, clip]{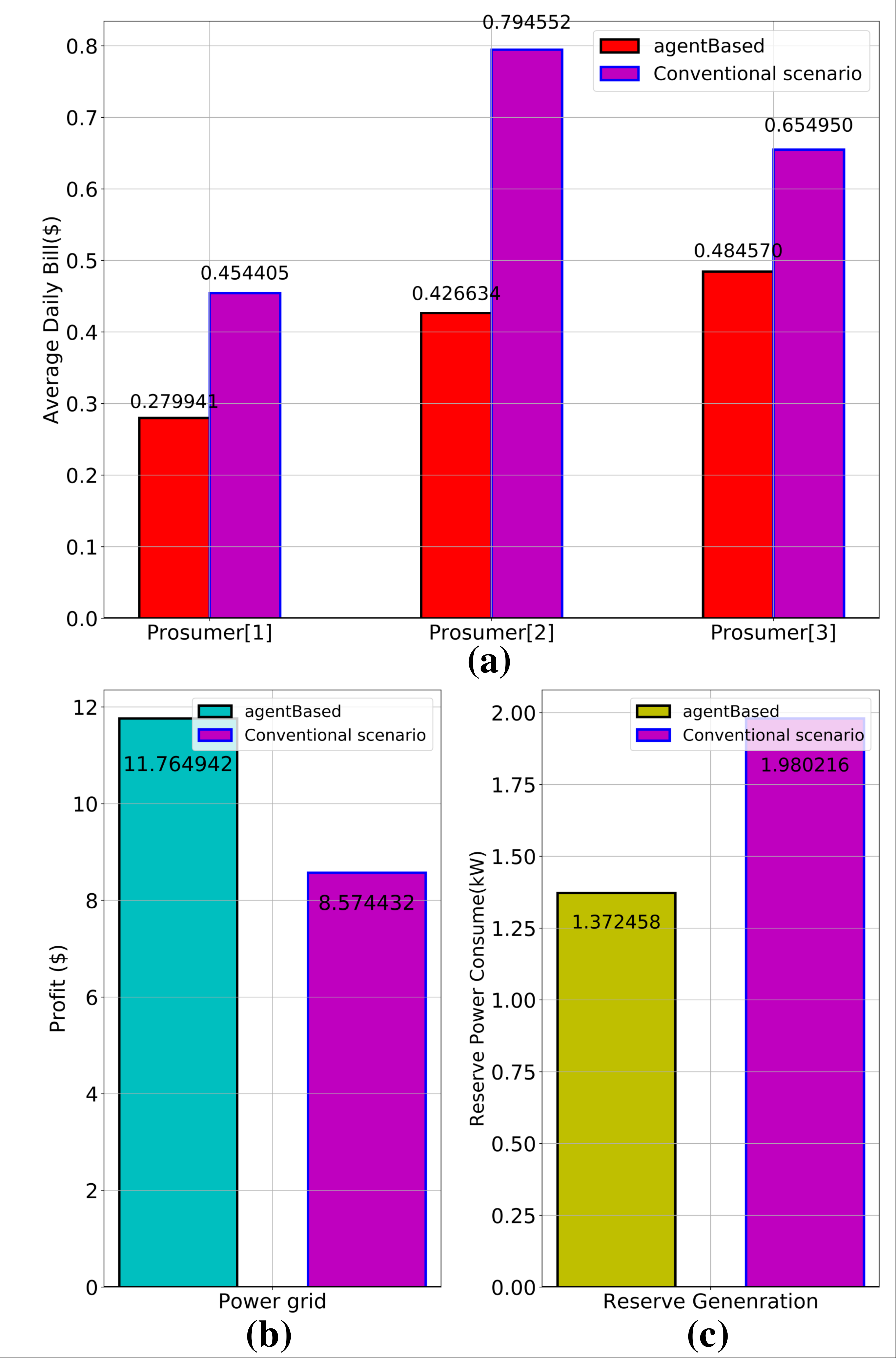}
    \caption{Daily bill comparison over episodes (a)prosumer 1-3 daily bill (b) grid profit (c) grid reserve power consume}
    \label{profit}
    \vspace{-.2in}
\end{figure}

Fig.~\ref{time-domain} compares the time domain profiles of the SoC of the prosumers and the buy/sell prices over a 24 hour period between the conventional and agent-based scenarios after fully training the DQN agents.  Comparing Fig.~\ref{time-domain} (b)-(d) with Fig.~\ref{waveform}~(a), it can be observed that in the conventional scenario the changes in the SoC of the battery of each prosumer is very closely synced with the prosumer's PV generation profile.  This is expected since as mentioned above in the conventional scenario prosumers inject power to grid when there is an excess energy generation.  On the other hand, in the agent based scenario, the battery SoC is changed as a result of charge/discharge commands issued by the  prosumer agents based on the identified optimal charge/discharge policy.  To compare the effect of this significant change in battery usage between the two scenarios, prosumers' average daily electricity bill, grid daily profit and reserve power consumption for the two scenarios have been calculated and plotted in  Fig.\ref{profit}.  According to this figure, the prosumers's average daily electricity bill is reduced significantly in the agent-based scenario.  Similarly, the grid profit is considerably higher in the agent-based scenario which can be attributed to the significant drop in the reserve generation usage in this scenario.                            

%The Buy price profile which is the action of the grid agent over a 24 hour period for maximizing grid side profit is shown in Fig.\ref{time-domain}(a).  Prosumers 1 to 3 conventional  and agent based battery SoC is shown in Fig.\ref{time-domain}(b)-(d). In the conventional scenario the battery tends to charge like PV generation trend which means the battery will be charged during the day time that there will be PV generation. On the other hand, in agent based scenario by participating the prosumers in DR scheme the prosumers motivated to use the full battery capacity in different times during 24 hours in order to reduce their electricity bill. Prosumer average daily bill, grid profit and reserve power consumption for two scenarios has been plotted in the Fig.\ref{profit} as  index to measure the effectiveness of this DR program. The daily bill reduction of 38\%, 46\% and 26\% respectively for prosumers 1, 2 and 3  from Fig.\ref{profit}(a). The power grid profit witnessed a 37\% increase in comparison to the conventional method in Fig.\ref{profit} (b). As such, the Fig.\ref{profit} (c) shows 30\% decrease in reserve generation usage.

According to the findings discussed above, it can be concluded that the grid and prosumer agents are leveraging the battery capacity of the prosumers to maximize their profits and reduce their costs.  

\begin{figure}[t]
   \centering
    \includegraphics[scale=.4]{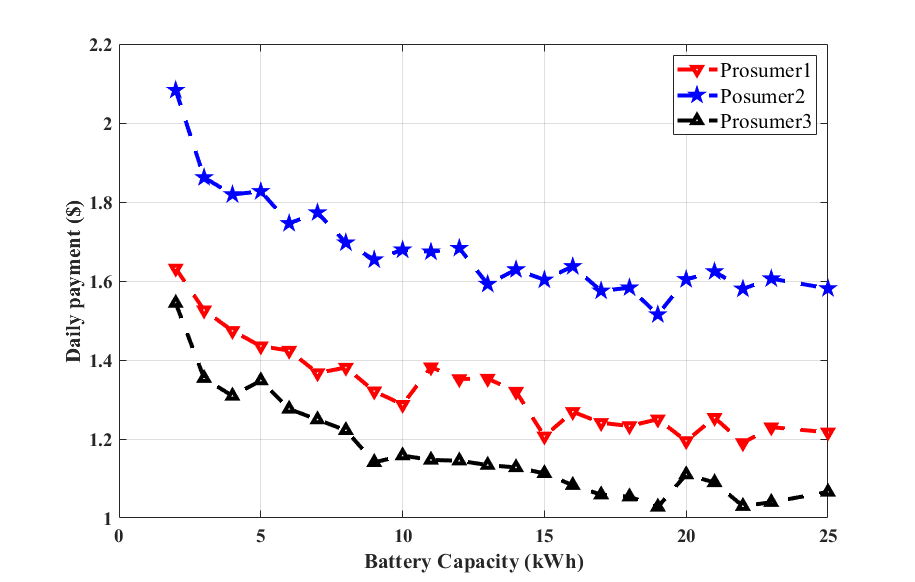}
    \caption{Daily bill reduction for prosumers with different battery capacity}
    \label{daily}
    \vspace{-.2in}
\end{figure}

\begin{figure}[t]
   \centering
    \includegraphics[scale=.4]{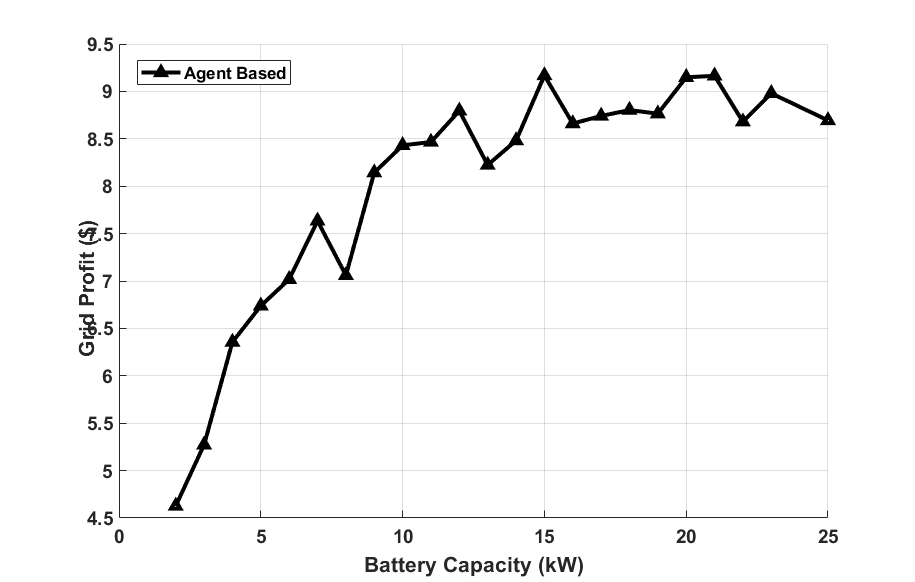}
    \caption{Monitoring Grid profit change with different battery capacity}
    \label{profit-capacity}
    \vspace{-.2in}
\end{figure}

In the next experiment, the effect of battery capacity on reduction of daily electricity bill of prosumers and raising the grid profit is investigated by running several agent-based simulations and increasing the battery capacity of the prosumers for each simulation.  The results are shown in Fig.~\ref{daily} and Fig.~\ref{profit-capacity}, respectively.  As pictured, the battery capacity is increased from 2 kWh to 25 kWh and the daily energy bill of prosumers and the grid profit are measured at the end of simualtion and plotted against the battery capacity.  Fig. \ref{daily} and Fig. \ref{profit-capacity}  show a downward trend in the daily electricity cost of prosumers and an upward trend in the grid profit as a function of battery capacity. However, these trends seem to slow down around 15 kWh battery capacity, meaning that the improvements when using batteries with larger than 15 kWh capacity seem negligible. Therefore, for a given PV generation capacity (i.e., $P^{pv,\max }$), it can be concluded that the proposed DR scheme can provide maximum benefits with a reasonable battery pack size of around 15 kWh in the households.

% The convergence of accumulative reward is illustrated in Fig~\ref{accumReward} for grid and prosumers. It can be seen from the the figure that in the beginning, the accumulative reward for both sides was low since the agents took inappropriate actions. Fig~\ref{accumReward} (a) and (b) depicts the moving average of accumulative reward for GA and three prosumers for 10000 Episodes with step 50. The decision making has become stronger and converged to the stable strategy as the agents explore more the environment by learning the optimal strategy.

% \begin{figure}
%   \centering
%     \includegraphics[scale=0.33]{SmartGridComm2020-paper/figs/accumulative-reward.pdf}
%     \caption{moving average of accumulative reward (a) grid reward (b) prosumers reward}
%     \label{accumReward}
%     \vspace{-.2in}
% \end{figure}

\section{Conclusions}~\label{sec:conclusions}
This paper proposes a new multiagent RL-based decision-making environment for implementing a DR scheme in a microgrid dominated by prosumers. The proposed technique implements a Real-Time Pricing scheme that can mitigate several shortcomings common to traditional DR methods while providing important economic benefits to the grid operator and prosumers.  
To showcase the better efficacy of the RL-based method, this work includes a comparison to a baseline traditional operation scenario in a small-scale microgrid system. Results showed significant daily bill reductions (e.g., 38\% , 46\%, and 26\%) for prosumers in the proposed RL-based marketplace . Moreover, experiments on the use of prosumers' energy storage capacity in this microgrid setup highlight the advantages of the proposed method in establishing a fair and balanced market setup. 
% The results showed 38\% , 46\% and 26\% daily bill reduction for prosumers 1 to 3 respectively in comparison with the conventional method.
% \mycomment{AB: It may be more convincing to briefly mention some numbers in the last statements. How much more efficient than the baseline? Perhaps also mention the existence of a threshold for battery capacity?}      

\bibliographystyle{IEEEtran}
\bibliography{ref}
\end{document}